\documentclass[graybox]{svmult}

\usepackage{mathptmx}       
\usepackage{helvet}         
\usepackage{courier}        
\usepackage{type1cm}        
\usepackage{makeidx}         
\usepackage{graphicx}        
\usepackage{multicol}        
\usepackage[bottom]{footmisc}

\usepackage{amsmath,amssymb}

\makeindex             

\renewcommand{\arraystretch}{1}

\begin{document}

\title*{Random Sampling: Practice Makes Imperfect}
\author{Philip B.~Stark and Kellie Ottoboni}
\institute{Philip B.~Stark \at University of California, Berkeley, \email{pbstark@berkeley.edu}
\and Kellie Ottoboni \at University of California, Berkeley \email{kellieotto@berkeley.edu}}
%
%
\maketitle

\abstract{
   The pseudo-random number generators (PRNGs), sampling algorithms, and algorithms for generating random integers in some common statistical packages and programming languages are unnecessarily
   inaccurate, by an amount that may matter for statistical inference.
    Most use PRNGs with state spaces that are too small for contemporary sampling problems and methods such as the bootstrap and permutation tests.
    The random sampling algorithms in many packages rely on the false assumption
    that PRNGs produce IID $U[0, 1)$ outputs.
    The discreteness of PRNG outputs and the limited state space of common PRNGs 
    cause those algorithms to perform poorly in practice.
    Statistics packages and scientific programming languages should use cryptographically secure PRNGs by default
    (not for their security properties, but for their statistical ones), and offer weaker PRNGs only as an option.
    Software should not use methods that assume PRNG outputs are IID $U[0,1)$ random variables, 
    such as generating a random sample by permuting the population and taking the first $k$ items 
    or generating random integers by multiplying a pseudo-random binary fraction or float by a constant and rounding the result.
    More accurate methods are available.
}


\begin{quotation}
The difference between theory and practice is smaller in theory than it is in practice. \\ ---Unknown
\end{quotation}

\begin{quotation}
In theory, there's no difference between theory and practice, but in practice, there is. \\ ---Jan L.A.~van~de~Snepscheut
\end{quotation}

\section{Introduction}
\label{sec:introduction}

Pseudo-random number generators (PRNGs) are central to the practice of Statistics.
They are used to draw random samples, allocate patients to treatments, perform the bootstrap, 
calibrate permutation tests, perform MCMC, approximate $p$-values, partition data into training and test sets, and countless other purposes.

Practitioners generally do not question whether standard software is adequate for these tasks.
This paper explores whether PRNGs generally considered adequate for statistical work really are
adequate, and whether standard software uses appropriate algorithms for generating random integers, random samples,
and independent and identically distributed (IID) random variates.

Textbooks give methods that implicitly or explicitly assume that PRNGs 
can be substituted for true IID $U[0,1)$ variables without introducing material error 
\cite{stine_statistics_2014, leblanc_statistics_2004, dahlberg_practical_2010, press_numerical_1988, peck_introduction_2011}.
We show here that this assumption is incorrect for algorithms in many commonly used statistical packages,
including MATLAB, Python's \texttt{random} module, R, SPSS, and Stata.

For example, whether software can in principle generate all samples of size $k$ from a population of
$n$ items---much less generate them with equal probability---depends on the size of the problem and the internals of the software, including the underlying PRNG and the \emph{sampling algorithm},
the method used to map PRNG output into a sample.
We show that even for datasets with hundreds of observations, 
many common PRNGs cannot draw all subsets of size $k$, for modest values of $k$.

Some sampling algorithms put greater demands on the PRNG than others.
For instance, some involve permuting the data.
The number of items that common PRNGs can permute ranges from at most 13 to at most 2084, far smaller than many data sets.
Other sampling algorithms require uniformly distributed integers (as opposed to the approximately $U[0,1)$ PRNG outputs) as input.
Many software packages generate
pseudo-random integers using a rounding method that does not yield uniformly distributed integers, even
if PRNG output were uniformly distributed on $w$-bit binary integers.

As a result of the limitations of common PRNGs and sampling algorithms,
the $L_1$ distance between the uniform distribution on samples of
size $k$ and the distribution induced by a particular PRNG and sampling algorithm can be nearly $2$.
It follows that there exist bounded functions of random samples whose expectations with respect to those two distributions differ substantially. 

Section~\ref{sec:prngs} presents an overview of PRNGs and gives examples of better and worse ones.
Section~\ref{sec:count} shows that, for modest $n$ and $k$, the state spaces of common PRNGs considered adequate for Statistics are too small to generate all permutations of 
$n$ things or all samples of $k$ of $n$ things. 
Section~\ref{sec:algorithms} discusses sampling algorithms and shows that some are less demanding on the PRNG than others. 
Section~\ref{sec:integers} shows that a common, ``textbook'' procedure for generating pseudo-random integers using a PRNG can be quite inaccurate; unfortunately, this is essentially the method that R uses
and that the Python \texttt{random.choice()} function uses.
Section~\ref{sec:discussion} concludes with recommendations and best practices.

\section{Pseudo-random number generators}
\label{sec:prngs}

A \emph{pseudo-random number generator} (PRNG) is a deterministic algorithm that, starting with
an initial ``seed'' value, produces a sequence of numbers that are supposed to behave like 
random numbers.
An ideal PRNG has output that is statistically indistinguishable from random, uniform, IID bits. 
Cryptographically secure PRNGs approach this ideal---the bits are (or seem to be) computationally 
indistinguishable from IID uniform bits---but common PRNGs do not.

A PRNG has several components:
an internal \emph{state}, initialized with a \emph{seed};
a function that maps the current state to an output;
and a function that updates the internal state.

If the state space is finite, the PRNG must eventually revisit a state after some number of calls---after 
which, it repeats.
The \emph{period} of a PRNG is the maximum, over initial states, of the number of states 
the PRNG visits before returning to a state already visited.
The period is at most the total number of possible states.
If the period is equal to the total number of states, the PRNG is said to have \emph{full period}.
PRNGs for which the state and the output are the same have periods no larger than the number of possible outputs.
Better PRNGs generally use a state space with dimension much larger than the dimension of the output.

Some PRNGs are sensitive to the initial state.
For unfavorable initial states, the PRNG may need many ``burn-in'' calls before the output behaves well.

\subsection{Simple PRNGs}


\emph{Linear congruential generators} (LCGs) have the form $X_{n+1} = (a X_n + c) \mod m$, for a 
\emph{modulus} $m$, \emph{multiplier} $a$, and \emph{additive constant} $c$.
LCGs are fast to compute and require little computer memory.
The behavior of LCGs is well understood from number theory.
For instance, the Hull-Dobell theorem \cite{hullDobell62}
gives necessary and sufficient conditions for a LCG to have full period for all seeds,
and there are upper bounds on the number of hyperplanes of dimension $k$
that contain all $k$-tuples of outputs, as a function of $m$ \cite{marsaglia_random_1968}.
When all $k$-tuples are in a smaller number of hyperplanes, that indicates that
the PRNG outputs are more regular and more predictable.

To take advantage of hardware efficiencies, early computer systems implemented LCGs 
with moduli of the form 
$m = 2^b$, where
$b$ was the integer word size of the computer.
This led to wide propagation of a particularly bad PRNG, RANDU, originally introduced
on IBM mainframes \cite{knuth_art_1997,markowsky14}.
(RANDU has $a=65539$, $m=2^{31}$, and $c=0$.)

More generally, LCGs with $m=2^b$ cannot have full period because $m$ is not prime. 
Better LCGs have been developed---and some are used in commercial statistical software packages---but they are still
generally considered inadequate for Statistics because of their short periods (typically $ \le 2^{32}$) 
and correlation among outputs.

The Wichmann-Hill PRNG is a sum of three normalized LCGs; its output is in $[0, 1)$.
%
It is generally not considered adequate for Statistics, but was (nominally) the PRNG in Excel 
2003, 2007, and 2010.\footnote{%
\url{https://support.microsoft.com/en-us/help/828795/description-of-the-rand-function-in-excel}, last visited 23 October 2018.}
The generator in Excel had an implementation bug that persisted for several generations.
Excel didn't allow the seed to be set so issues could not be replicated, but users reported that the PRNG occasionally gave a negative output \cite{mccullough_microsoft_2008}.
As of 2014, IMF banking Stress tests used Excel simulations \cite{ong14}.
This worries us.

Many other approaches to generating pseudo-random numbers have been
proposed, and PRNGs can be built by combining simpler ones (carefully---see \cite{knuth_art_1997} on ``randomly'' combining PRNGs).
For instance, the KISS generator combines four generators of three types, and has a period greater than $2^{210}$.
Nonetheless, standard PRNGs are predictable from a relatively small number of outputs.
For example, one can determine the LCG constants $a$, $c$, and $m$ by observing only 3 outputs,
and can recover the state of KISS from about 70~words of output \cite{rose11}.

\subsection{Mersenne Twister (MT)}

Mersenne Twister (MT) \cite{matsumoto_mersenne_1998} is a ``twisted generalized feedback shift register,'' a sequence of bitwise and linear operations.
Its state space is $19,937$ bits and it has an enormous period $2^{19937}-1$, a Mersenne prime.
It is $k$-equidistributed to $32$-bit accuracy for $k \leq 623$, 
meaning that output vectors of length up to $623$ (except the zero vector) occur with equal frequency over the full period.
The state is a $624 \times 32$ binary matrix.

MT is the default PRNG in common languages and software packages, including Python, R, Stata, 
GNU Octave, Maple, MATLAB, Mathematica, and many more (see Table~\ref{tab:software}).
We show below that it is not adequate for statistical analysis of modern data sets.
Moreover, MT can have slow ``burn in,'' especially for seeds with many zeros \cite{saito_simd-oriented_2008}.
And the outputs for close seeds can be similar, which makes seeding distributed computations
delicate.

\subsection{Cryptographic hash functions}
The PRNGs described above are quick to compute but predictable,
and their outputs are easy to distinguish from actual random bits \cite{lecuyer_testu01_2007}.
Cryptographers have devoted a great deal of energy to inventing cryptographic hash functions,
which can be used to create PRNGs, as the properties that make functions cryptographically secure
are properties of good pseudo-randomness.

A \emph{cryptographic hash function} $H$ is a function with the following properties:

\begin{itemize}
\item $H$ produces a fixed-length ``digest'' (hash) from arbitrarily long ``message'' (input):\\ $H:\{0, 1\}^* \rightarrow \{0, 1\}^L$.
\item $H$ is inexpensive to compute.
\item $H$ is ``one-way,'' i.e., it is hard to find a pre-image of any output except by exhaustive enumeration (this is the basis of  hashcash ``proof of work'' for Bitcoin and some other distributed ledgers).
\item $H$ is collision-resistant, i.e., it is hard to find $M_1 \ne M_2$ such that $H(M_1) = H(M_2)$.
\item small changes to $M$ produce unpredictable, big changes to $H(M)$.
\item outputs of $H$ are equidistributed: bits of the hash are essentially IID random.
\end{itemize}

These properties of $H$ make it suitable as the basis of a PRNG:
It is \emph{as if} $H(M)$ is a uniformly distributed random $L$-bit string assigned to $M$.
One can construct a simple hash-based PRNG with the following procedure, which we first learned about
from Ronald L.~Rivest:

\begin{enumerate}
\item Generate a random string $S$ with a substantial amount of entropy, e.g., 20 rolls of a
10-sided die.
\item Set $i=0$. The state of the PRNG is the string ``S,i''. $i$ is the number of values generated so far. 
\item Set $X_i = {\mbox{Hash}}(S,i)$, interpreted as a (long) hexadecimal number.
\item Increment $i$ and return to step~3 to generate more outputs.
\end{enumerate}

\noindent Since a message can be arbitrarily long, this PRNG has an unbounded state space. For truly cryptographic applications, the seed should be reset to a new random value periodically; for statistical applications, that should not be necessary.

\section{Counting permutations and samples}
\label{sec:count}

\begin{theorem}[Pigeonhole principle]
If you put $N>n$ pigeons in $n$ pigeonholes, at least one
pigeonhole must contain more than one pigeon.
\end{theorem}

\begin{corollary}
At most $n$ pigeons can be put in $n$ pigeonholes if at most
one pigeon is put in each hole.
\end{corollary}

The corollary implies that a PRNG cannot generate more permutations or samples than the number of states the PRNG has (which is in turn an upper bound on the period of the PRNG).
Of course, that does not mean that the permutations or samples a PRNG can generate occur with approximately equal probability: that depends on the quality of the PRNG, not just the size of the state space. 
Nonetheless, it follows that no PRNG with a finite state space can be ``adequate for Statistics'' for every statistical problem.

The number of permutations of $n$ objects is $n!$, the number of possible samples of $k$ of $n$ items
with replacement is $n^k$,  and the number of possible samples of $k$ of $n$ without replacement is $\tbinom{n}{k}$.
These bounds are helpful for counting pigeons:
\begin{itemize}
\item Stirling bounds: $ e n^{n+1/2} e^{-n} \ge n! \ge \sqrt{2 \pi} n^{n+1/2} e^{-n}.$
\item Entropy bounds:
$ \frac{2^{nH(k/n)}}{n+1} \le \tbinom{n}{k} \le 2^{nH(k/n)},$ where $H(q) \equiv -q \log_2(q) - (1-q) \log_2 (1-q)$.
\item Stirling combination bounds:
for $\ell \ge 1$ and $m \ge 2$, $ \tbinom{\ell m}{\ell } \ge \frac{m^{m(\ell-1)+1}}{\sqrt{\ell} (m-1)^{(m-1)(\ell-1)}}. $
\end{itemize}

Table~\ref{tab:pigeonhole} compares numbers of permutations and random samples to the size of
the state space of various PRNGs.
PRNGs with 32-bit state spaces, which include some in statistical packages, cannot generate all permutations of even small populations, nor all random samples of small size from modest populations.
MT is better, but still inadequate: it can generate fewer than 1\% of the permutations of 2084 items.

\begin{table}
\caption{The pigeonhole principle applied to PRNGs, samples, and permutations.
For a PRNG of each size state space, the table gives examples where some samples or permutations 
must be unobtainable.}
\label{tab:pigeonhole}       
\begin{tabular}[h]{p{4.7cm}p{2.4cm}p{3.9cm}p{2cm}}
\hline\noalign{\smallskip}
Feature & Size & Full & Scientific \\
            &        &       & notation  \\
\noalign{\smallskip}\svhline\noalign{\smallskip}
32-bit state space & $2^{32}$ & 4,294,967,296 & $4.29 \times 10^9$ \\
Permutations of 13 & $13!$ & 6,227,020,800 & $6.23 \times 10^9$ \\
Samples of 10 out of 50 & $\tbinom{50}{10}$ &  10,272,278,170 & $1.03\times 10^{10} $ \\
{}\\
Fraction of attainable samples with 32-bit state space & ${2^{32}}/{\tbinom{50}{10}}$   & 0.418 & \\
\noalign{\smallskip}\svhline\noalign{\smallskip}
64-bit state space & $2^{64}$ & 18,446,744,073,709,551,616 & $1.84 \times 10^{19}$ \\
Permutations of 21 & $21!$ &  51,090,942,171,709,440,000 & $5.11 \times 10^{19}$ \\
Samples of 10 out of 500 & $\tbinom{500}{10}$ & & $2.46 \times 10^{20}$ \\
{}\\
Fraction of attainable samples with 64-bit state space & ${2^{64}}/{\tbinom{500}{10}}$ &  0.075 & \\
\noalign{\smallskip}\svhline\noalign{\smallskip}
128-bit state space & $2^{128}$ &  & $3.40 \times 10^{38}$ \\
Permutations of 35 & $35!$ &   & $1.03 \times 10^{40}$ \\
Samples of 25 out of 500 & $\tbinom{500}{25}$ & & $2.67 \times 10^{42}$ \\
{}\\
Fraction of attainable samples with 128-bit state space & ${2^{128}}/{\tbinom{500}{25}}$ &  0.0003 & \\
\noalign{\smallskip}\svhline\noalign{\smallskip}
MT state space & $2^{32 \times 624}$ & & $9.27\times 10^{6010}$ \\
Permutations of 2084 & $2084!$ &   & $3.73 \times 10^{6013}$ \\
Samples of 1000 out of 390 million & $\tbinom{3.9\times 10^8}{1000}$ & & $> 10^{6016}$ \\
Fraction of attainable samples & ${2^{32 \times 624}}/{\tbinom{3.9\times 10^8}{1000}}$ &  & $< 1.66 \times 10^{-6}$ \\
\noalign{\smallskip}\svhline\noalign{\smallskip}
\end{tabular}
\end{table}

\subsection{$L_1$ bounds}\label{sec:L1bounds}

Simple probability inequalities give attainable bounds on the bias introduced by using a 
PRNG with insufficiently large state space, on  the assumption that the PRNG is uniform on its possible outputs. 
(Failure of that uniformity makes matters even worse.)
Suppose ${\mathbb P}_0$ and ${\mathbb P}_1$ are probability distributions on a common measurable space. 
If there is some measurable set $S$ for which ${\mathbb P}_0(S) = \epsilon$ and ${\mathbb P}_1(S) = 0$, then $\|{\mathbb P}_0 - {\mathbb P}_1 \|_1 \ge 2 \epsilon$.
Thus there is a function $f$ with $|f| \le 1$ such that 

$${\mathbb E}_{{\mathbb P}_0}f -  {\mathbb E}_{{\mathbb P}_1}f \ge 2 \epsilon.$$

In the present context, ${\mathbb P}_0$ is the uniform distribution (on samples or permutations)
and ${\mathbb P}_1$ is the distribution induced by the PRNG and sampling algorithm.
If the PRNG has $n$ states and we want to generate $N>n$ equally likely outcomes, at least $N-n$ outcomes will have probability zero instead of $1/N$.
Some statistics will have bias of (at least) $2 \times \frac{N-n}{N}$.
As seen in Table~\ref{tab:pigeonhole}, the fraction of attainable samples or 
permutations is quite small in problems of a size commonly encountered
in practice, making the bias nearly $2$. 

\section{Sampling algorithms}
\label{sec:algorithms}

There are many ways to use a source of pseudo-randomness to simulate drawing a simple random sample. 
A common approach is like shuffling a deck of $n$ cards, then dealing the top $k$:
assign a (pseudo-)random number to each item, sort the items based on that number
to produce a permutation of the population, 
then take the first $k$ elements of the permuted list to be the sample
\cite{stine_statistics_2014, leblanc_statistics_2004, dahlberg_practical_2010}.
We call this algorithm PIKK: permute indices and keep $k$.

If the pseudo-random numbers really were IID $U[0,1)$,  every permutation would indeed be equally likely, 
and the first $k$ would be a simple random sample.
But if the permutations are not equiprobable, there is no reason to think that the first $k$ 
elements comprise a random sample. 
Furthermore, this algorithm is inefficient: it requires generating 
$n$ pseudo-random numbers and then an $O(n\log n)$ sorting operation.

There are better ways to generate a random permutation, such as the ``Fisher-Yates shuffle'' or ``Knuth shuffle'' (Knuth attributes it to Durstenfeld) \cite{knuth_art_1997},
which involves generating $n$ independent random integers on various ranges, but no sorting.
There is also a version suitable for \emph{streaming}, i.e., permuting a list that has an (initially) unknown number of elements.
Generating $n$ pseudo-random numbers places more demand on a PRNG than other sampling algorithms discussed below, which only require $k<n$ pseudo-random numbers.

One simple method to draw a random sample of size $k$ from a population of size $n$
is to draw $k$ integers at random without replacement from  $\{1, \ldots, n\}$, then take the items with those indices to be the sample.
\cite{cormen_introduction_2009} provide an elegant recursive algorithm to draw random samples of size $k$ out of $n$; it requires the software recursion limit to be at least $k$.
(In Python, the default maximum recursion depth is $2000$, so this algorithm cannot draw samples of size greater than $2000$ unless one increases the recursion limit.)

The sampling algorithms mentioned so far require $n$ to be known.
\emph{Reservoir} algorithms, such as Waterman's Algorithm $R$, do not  \cite{knuth_art_1997}.
Moreover, reservoir algorithms are suitable for streaming: items are examined
sequentially and either enter into the reservoir, or, if not, are never revisited.
Vitter's Algorithm $Z$ is even more efficient than Algorithm $R$,
using random skips to reduce runtime to be essentially linear in $k$ \cite{vitter_random_1985}.

\subsection{Pseudo-random integers}
\label{sec:integers}
Many sampling algorithms require pseudo-random integers on $\{1, \ldots, m\}$.
The output of a PRNG is typically a $w$-bit integer, so some method is needed to map it to the range $\{1, \ldots, m\}$.

A textbook way to generate an integer on the range $\{1, \ldots, m\}$ is to first draw a random $X \sim U[0,1)$
and then define $Y \equiv 1 + \lfloor mX \rfloor$ \cite{press_numerical_1988, peck_introduction_2011}. 
In practice, PRNG outputs are not $U[0,1)$: they are derived by normalizing a value that is
 (supposed to be) uniformly distributed on $w$-bit integers. 

Even if $X$ is uniformly distributed on $w$-bit integers, the distribution of $Y$ will not be uniform on $\{1, \ldots, m\}$ unless $m$ is a power of 2.
If $m > 2^w$, at least $m-2^w$ values will have probability 0 instead of probability $1/m$.
If $w=32$, then for $m>2^{32}\approx4.24 \times 10^9$, some values will have probability 0. 
Conversely, there exists $m < 2^w$ such that the ratio of the largest to smallest selection probability
of $\{1, \ldots, m\}$ is, to first order,  $1+ m 2^{-w+1}$ \cite{knuth_art_1997}.

R (Version 3.5.1) \cite{R_2018} uses this multiply-and-floor approach to generate 
pseudo-random integers,
which eventually are used in the main sampling functions.
Duncan Murdoch devised a simple simulation that shows how large the inhomogeneity of selection
probabilities can be:
for $m=  (2/5) \times 2^{32} = 1,717,986,918$, the \texttt{sample()} function generates about 40\% even numbers and about 60\% odd numbers \footnote{ %
\url{https://stat.ethz.ch/pipermail/r-devel/2018-September/076827.html}, last visited 17 October 2018
}. %
    
A more accurate way to generate random integers on $\{1, \dots, m\}$ is to use pseudo-random bits directly. 
This is not a new idea; \cite{hodges_basic_1970} describe essentially the same procedure to draw integers by hand
from random decimal digit tables.
The integer $m-1$ can be represented with $\mu = \lceil \log_2(m-1) \rceil$ bits. 
To generate a pseudo-random integer uniformly distributed on $\{1, \ldots, m\}$, 
generate $\mu$ pseudo-random bits (for instance, by taking the most significant $\mu$ bits from the PRNG output) and interpret the bits as a binary integer.  
If the integer is larger than $m-1$, then discard it and draw another $\mu$ bits until the $\mu$ bits represent an integer less than or equal to $m-1$.
When that occurs, return that integer, plus 1.
This procedure potentially requires throwing out (in expectation) almost half the draws if $m-1$ is 
just below a power of $2$, but the algorithm's output will be uniformly distributed (if the input bits are).
This is how the Python package Numpy (Version 1.14) generates pseudo-random integers.\footnote{
However, Python's built-in \texttt{random.choice()} (Versions 2.7 through 3.6) does something else that's biased: it finds the closest integer to $mX$.
}

\section{Discussion}
\label{sec:discussion}

Any PRNG with a finite state space cannot generate all possible samples from or permutations of 
sufficiently large populations.
That can matter.
A PRNG with a 32-bit state space cannot generate all permutations of 13 items.
MT cannot generate all permutations of 2084 items.

Table~\ref{tab:software} lists the PRNGs and sampling algorithms used in common statistical packages.
Most use MT as their default PRNG; \emph{is} MT adequate for Statistics?
Section~\ref{sec:L1bounds} shows that for some statistics, the $L_1$ distance between the theoretical value and the attainable value using a given PRNG
 is big for even modest sampling and permutation problems.
We have been searching for biases that are large enough to matter in 
$O(10^5)$ replications or less, and are not idiosyncratic to a few bad seeds.
We have examined the frequencies of simple random samples, the frequency of derangements and partial derangements, the Spearman correlation between permutations, 
and other statistics; so far, we have not found a statistic with consistent
bias large enough to be detected in $O(10^5)$ replications.
MT must produce bias in some statistics, but which?

\renewcommand{\arraystretch}{1}

\begin{table}
\caption{PRNGs and sampling algorithms used in common statistical and mathematical software packages. The `floor' algorithm is the flawed multiply-and-floor method of generating pseudo-random integers. The `mask' algorithm is better.}
\label{tab:software}      
\begin{tabular}[h]{p{2.5cm}p{2.2cm}p{1.8cm}p{4.9cm}}
\hline\noalign{\smallskip}
Package/Language & Default PRNG & Other & SRS Algorithm  \\
\noalign{\smallskip}\svhline\noalign{\smallskip}
SAS 9.2              & MT         	& 32-bit LCG & Floyd's ordered hash or Fan's method \cite{fan_development_1962} \\
SPSS 20.0          & 32-bit LCG  & MT1997ar  & floor + random indices \\
SPSS $\le$ 12.0 & 32-bit LCG  &         &                \\
STATA 13            & KISS 32      &         & PIKK           \\
STATA 14            & MT              &         & PIKK           \\
R                         & MT              &         & floor + random indices \\
Python                 & MT             &         & mask + random indices  \\
MATLAB              & MT             &         & floor + PIKK         \\
\noalign{\smallskip}\hline\noalign{\smallskip}
\end{tabular}
\end{table}

We recommend the following practices and considerations for using PRNGs in Statistics:
\begin{itemize}
\item Consider the size of the problem: are your PRNG and sampling algorithm adequate?
\item Use a source of real randomness to set the seed with a substantial amount of entropy, e.g., 20 rolls of 10-sided dice.
\item Record the seed so your analysis is reproducible.
\item Avoid standard linear congruential generators, the Wichmann-Hill generator, and PRNGs with small state spaces.
\item Use a cryptographically secure PRNG unless you know that MT is adequate for your problem.
\item Use a sampling algorithm that does not overtax the PRNG. Avoid permuting the entire population to draw a random sample: do not use PIKK.
\item Beware discretization issues in the sampling algorithm; many methods assume the PRNG produces $U[0,1]$ or $U[0,1)$ random numbers, rather than (an approximation to) numbers that are uniform on $w$-bit binary integers.
\end{itemize}

We also recommend that R and Python upgrade their algorithms to use best practices.
R should replace the multiply-and-floor algorithm it uses to generate random integers in the \texttt{sample} function (and other functions) with the more precise bit masking algorithm, as discussed
in \cite{ottoboniStark18}.
And we suggest R and Python use cryptographically secure PRNGs by default, with an option of using MT instead in case the difference in speed matters.
We have developed a CS-PRNG prototype as a Python package, \texttt{cryptorandom}.\footnote{ %
\texttt{cryptorandom} can be downloaded at \url{https://github.com/statlab/cryptorandom} and \url{https://pypi.org/project/cryptorandom/}.
} %
The current implementation is slow (the bottleneck is Python data type conversions, not computation);
we are developing a faster C implementation.

\begin{acknowledgement}
We are grateful to Ronald L.~Rivest for many helpful conversations.
\end{acknowledgement}

\bibliographystyle{spmpsci}
\bibliography{refs}

\end{document}